\documentstyle[11pt,a4,axodraw]{article}

\newcommand{\nc}[2]{\newcommand{#1}{#2}}
\newcommand{\ncx}[3]{\newcommand{#1}[#2]{#3}}
\ncx{\pr}{1}{#1^{\prime}}
\nc{\nl}{\newline}
\nc{\np}{\newpage}
\nc{\nit}{\noindent}
\nc{\be}{\begin{equation}}
\nc{\ee}{\end{equation}}
\nc{\ba}{\begin{array}}
\nc{\ea}{\end{array}}
\nc{\dsp}{\displaystyle}
\nc{\bit}{\bibitem}
\nc{\ct}{\cite}
\ncx{\dd}{2}{\frac{\partial #1}{\partial #2}}
\nc{\pl}{\partial}
\nc{\dg}{\dagger}
\nc{\ag}{\alpha}
\nc{\bg}{\beta}
\nc{\gam}{\gamma}
\nc{\Gam}{\Gamma}
\nc{\bgm}{\bar{\gam}}
\nc{\del}{\delta}
\nc{\Del}{\Delta}
\nc{\eps}{\epsilon}
\nc{\ve}{\varepsilon}
\nc{\kg}{\kappa}
\nc{\lb}{\lambda}
\nc{\ps}{\psi}
\nc{\Ps}{\Psi}
\nc{\sg}{\sigma}
\nc{\spr}{\pr{\sg}}
\nc{\Sg}{\Sigma}
\nc{\rg}{\rho}
\nc{\fg}{\phi}
\nc{\Fg}{\Phi}
\nc{\vf}{\varphi}
\nc{\og}{\omega}
\nc{\Og}{\Omega}
\nc{\Ztwo}{\mbox{\sf Z}_{2}}
\nc{\Tr}{\mbox{Tr}}
\nc{\lh}{\left(}
\nc{\rh}{\right)}
\nc{\cB}{\mbox{$^{\ast}\Og$ }}
\nc{\nil}{\emptyset}
\nc{\bor}{\overline}

\begin{document}

\pagestyle{empty}

\begin{flushright}
NIKHEF-H/94-22
\end{flushright}

\begin{center}

{\Large {\bf Relations between some analytic representations }} \\
\vspace{3ex}

{\Large {\bf of one-loop scalar integrals}} \\
\vspace{5ex}

{\large J.W. van Holten} \\
\vspace{3ex}

{\large NIKHEF-H, Amsterdam NL} \\
\vspace{5ex}

August 1, 1994
\vspace{7ex}

{\small {\bf Abstract}}

\end{center}

\nit
{\small
We compare several parametrized analytic expressions for an arbitrary off-shell
one-loop $n$-point function in scalar field theory in $D$-dimensional
space-time, and show their equivalence both directly and through path-integral
methods.}

\np

\pagestyle{plain}
\pagenumbering{arabic}

\section{Introduction}{\label{S1}}

Inspired by the string-based work of Bern and Kosower \ct{BK1,BK2}, Strassler
\ct{MS} has given a set of rules for computing one-loop Green functions in
field theory directly from path-integrals for relativistic point-particle
models. Similar results in this direction were already obtained earlier by
Polyakov \ct{Pol}. The procedure has been extended by various authors
\ct{McK}-\ct{CS3} so as to allow for example the computation of processes with
internal or external fermions, of higher-order terms in the
derivative-expansion of the effective potential, or of higher-loops. In many
of these applications, the emphasis is on the non-trivial structure due to
spin-polarization and non-abelian charges, and in particular on the manifest
gauge-invariance of the computational scheme, whilst the scalar loop-integrals
are merely shown to reduce to a standard form. An exception is reference
\ct{CS2}, where use of the method is made to construct the derivative expansion
of the effective action of scalar field theory.

It is the purpose of this note to show how the expressions for the scalar
loop-integrals, first obtained by the path-integral method, can also be derived
directly from the standard Feynman-diagram representation. This provides an
independent check of the correctness of the path-integral manipulations or,
if one prefers, of the equivalence between the Feynman-diagram method and
the functional-integral representation of the corresponding point-particle
theory. Indeed, a step-by-step comparison between the path-integral and Feynman
diagram calculation becomes possible. Basically it is another way of
establishing the equivalence between `first' and `second' quantization.

\section{The one-loop scalar $n$-point function}{\label{S2}}

Consider the one-loop diagrams for $g \vf^N$-theory with mass $m$ in
$D$-dimensional Minkowi\-ski space, see fig.\ 1. Since the particles in the
theory are structureless, the only effect of any vertex in the loop is a
multiplicative factor $-g /(2 \pi)^D$, an additional propagator, and the
insertion of a momentum

\be
Q^{\mu} = \sum_{r=1}^{N-2} \, k_r^{\mu},
\label{1}
\ee

\nit
into the loop; here the $k_r^{\mu}$ are the (off-shell) momenta of the $N-2$
individual external lines entering the vertex. The full one-loop contribution
to
the $n$-point function is obtained by symmetrization over all vertices, or
equivalently all external total momenta $Q_i$, $i = 1,...,n$, and taking into
account over-all momentum conservation:

\be
\sum_{i=1}^{n}\, Q_i = 0.
\label{1.1}
\ee

\thicklines

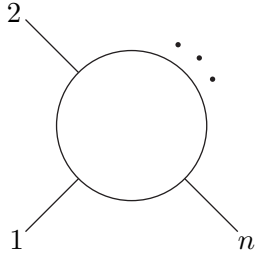
\begin{figure}
\begin{picture}(150,150)(-130,10)


\BCirc(75,75){28.3}

\Line(55,55)(35,35)
\Text(29,32)[l]{1}
\Line(55,95)(35,115)
\Text(28,118)[l]{2}
\Line(95,55)(115,35)
\Text(122,31)[r]{$n$}

\Vertex(100.5,100.5){1}
\Vertex(105.5,92.5){1}
\Vertex(92.5,105.5){1}

\end{picture}
\caption{1-loop scalar $n$-point diagram}
\end{figure}

\nit
The well-known result for the amplitude is

\be
\ba{lll}
\Gam_n [Q_i] & = & \dsp{ \frac{1}{n}\, \left[ \frac{-g}{(2\pi)^D} \right]^n \,
           \del^{(D)} \lh \sum_{k=1}^{n} Q_k \rh\, \times }\\
  &  & \\
 & & \dsp{ \times \sum_{\left\{ i_k \right\}} \int_0^1 \prod_{i=1}^{n} dx_i\,
           \int d^D p \frac{\del \lh 1 - \sum_{j=1}^{n} x_j \rh }{\left[
           \sum_{l=1}^{n} x_l p_l^2\, +\, m^2\, -\, i \eps \right]^n }, }
\label{2}
\ea
\ee

\nit
where $\left\{i_k\right\}$ denotes a permutation of $(1,...,n)$, whilst

\be
p_1 = p,  \hspace{2em} \mbox{and} \hspace{2em}
p_l = p + \sum_{k=1}^{l-1} Q_{i_k}, \hspace{1em} \mbox{for} \hspace{1em}
      l = 2,...,n.
\label{3}
\ee

\nit
Clearly the last expression depends on the particular permutation of the $Q_i$.
The $x_i$ with values in the interval $[0,1]$ are the usual Feynman parameters.

The integrand in eq.(\ref{2}) can be converted to exponential form using the
Schwinger trick; introducing a factor $1/2m$ for later convenience, we write

\be
\frac{1}{\left[ \sum_{j=1}^{n} x_j p_j^2\, +\, m^2\, -\, i \eps \right]^n }\,
  =\, \frac{i^n}{(n-1)! (2m)^n}\, \int_0^{\infty} dT\, T^{n-1}
      e^{-\frac{iT}{2m}\, \sum_{j=1}^{n}\, x_j (p_j^2 + m^2 - i\eps)}.
\label{4}
\ee

\nit
The factor $T$ can be removed from the exponent by absorbing it into the
Feynman
parameters; hence we define

\be
\ag_i \equiv T x_i,
\label{5}
\ee

\nit
taking values in $[0,T]$. Thus we obtain the following expression
for the $n$-point amplitude in one loop:

\be
\ba{lll}
\Gam_n [Q_i] & = & \dsp{
   \frac{1}{n!}\, \left[ \frac{-i g}{2m (2\pi)^D} \right]^n\,
   \del^{(D)} \lh \sum_{k=1}^{n} Q_k \rh \times }\\
  & & \\
  & & \dsp{ \times \sum_{\left\{ i_k \right\} }\, \int_0^{\infty} dT\,
      \int_0^T \prod_{i=1}^{n} d\ag_i \, \del \lh T - \sum_{i=1}^{n} \ag_i
      \rh\, \int d^D p e^{- \frac{i}{2m}\, \sum_{j=1}^{n}\, \ag_j (p^2_j +
       m^2)}.}
\ea
\label{6}
\ee

\nit
In the exponent we have tacitly absorbed the $i\eps$ term in the mass $m^2$,
and
we will no longer write it explicitly.

The integral over the loop-momentum $p$ has now been reduced to a Gaussian
and is therefore straightforward to perform. One finds (in Minkowski space):

\be
\ba{lll}
\Gam_n [Q_i] & = & \dsp{
   \frac{i}{n!}\, \left[ \frac{-i g}{2m (2\pi)^D} \right]^n\,
   \del^D \lh \sum_{k=1}^{n} Q_k \rh \times }\\
  &  &  \\
  &  & \dsp{ \times \sum_{\left\{ i_k \right\} }\, \int_0^{\infty} dT\,
       \int_0^T \prod_{i=1}^{n} d\ag_i \, \del \lh T - \sum_{i=1}^{n} \ag_i
       \rh\, \lh \frac{-2 \pi i m }{T} \rh^{D/2}\, e^{-\, \frac{imT}{2}\, -\,
       \frac{i}{2mT}\, R_n \left[ Q_i \right] },}
\ea
\label{7}
\ee

\nit
where $R_n \left[ Q_i \right]$ is a quadratic polynomial in the external
momenta; explicitly:

\be
R_n \left[ Q_i \right]\, =\, T\, \sum_{j=2}^{n}\, \left[ \ag_j
    \lh \sum_{k=1}^{j-1}\, Q_{i_k} \rh^2 \right]\, -\,
    \sum_{j=2}^{n}\, \left[ \ag_j \sum_{k=1}^{j-1} Q_{i_k} \right]^2.
\label{8}
\ee

\nit
The sums on the right-hand side start only at $j = 2$, because $\ag_1$ is
multiplied by $\sum_{k=1}^{n} Q_{i_k}$, which vanishes acording to
eq.(\ref{1.1}) in any permutation.

We proceed to show that in the symmetric integral over the parameters $\ag_i$
the polynomial $R_n \left[ Q_i \right]$ may be replaced by

\be
R_n \left[ Q_i \right]\, =\, 2 T\, \sum_{k=2}^{n} \sum_{l=1}^{k-1}\,
    Q_{i_k} \cdot Q_{i_l}\, G\left[ \Del_{kl} \right],
\label{9}
\ee

\nit
where the function $G[\Del_{kl}]$ of the argument $\Del_{kl} =
\sum_{j=l+1}^{k}\,
\ag_j$ is defined as

\be
G[\Del_{kl}]\, =\, -\frac{1}{2}\, \left| \Del_{kl} \right|\, +\,
                   \frac{\Del_{kl}^2}{2T}.
\label{10}
\ee

\nit
This rearrangement of terms is possible, because of the momentum conservation
condition (\ref{1.1}) and the constraint

\be
\sum_{i=1}^{n}\, \ag_i = T,
\label{11}
\ee

\nit
which show the full set of variables and parameters in the polynomial $R_n
\left[ Q_i \right]$ to be redundant. Since the proof is independend of the
ordering of the external momenta, we may take any particular ordering. For
convenience of notation we take $\left\{ Q_{i_k} \right\} = \left\{ Q_{k}
\right\}$. Using the resummation formula

\be
\sum_{k=2}^{n}\, \lh \ag_k \sum_{l=1}^{k-1}\, B_l \rh\, =\,
      \sum_{l=1}^{n-1}\, \lh B_l \sum_{k=l+1}^{n}\, \ag_k \rh,
\label{12}
\ee

\nit
and splitting off the terms which contain pure squares:

\be
\sum_{k,l = 1}^{n-1}\, c_{kl}\, Q_k \cdot Q_l\, =\,
  \sum_{k=1}^{n-1}\, c_{kk}\, Q_k^2\, +\,
  2 \sum_{k=2}^{n-1} \sum_{l=1}^{k-1}\, c_{kl}\, Q_k \cdot Q_l,
\label{13}
\ee

\nit
we can rewrite the expression (\ref{8}) for $R_n \left[ Q_i \right]$ in the
form

\be
\ba{lll}
R_n \left[ Q_i \right] & = & \dsp{ \sum_{k=1}^{n-1}\, Q_k^2 \left[
    T \sum_{j=k+1}^n \ag_j\, -\, \lh \sum_{j=k+1}^n \ag_j \rh^2 \right]\, + }\\
 & & \\
 & & \dsp{  +\, 2\, \sum_{k=2}^{n-1} \sum_{l=1}^{k-1}\, Q_k \cdot Q_l \left[
    T \sum_{j=k+1}^n \ag_j\, -\, \lh \sum_{j=k+1}^n \ag_j \rh^2 \right]. }
\label{14}
\ea
\ee

\nit
Note that the mixed term has been arranged such that the sum runs only over
$k>l$,
giving rise to the factor of 2. Now we can eliminate the pure squares by using
momentum conservation:

\be
\ba{lll}
Q_1^2 & = & \dsp{ - Q_1 \cdot \sum_{l=2}^{n} Q_l , } \\
  & & \\
Q_k^2 & = & \dsp{ - Q_k \cdot \lh \sum_{l=1}^{k-1} Q_l + \sum_{l=k+1}^n Q_l
\rh,
\hspace{3em} k \geq 2,}
\label{15}
\ea
\ee

\nit
Using the constraint (\ref{11}) to expand $T$ in terms of the $\ag_i$,
eq.(\ref{14}) for $R_n \left[ Q_i \right]$ then becomes

\be
\ba{lll}
R_n \left[ Q_i \right] & = & \dsp{
  \sum_{k=2}^{n-1} \sum_{l=1}^{k-1}\, Q_k \cdot Q_l
  \left[ 2 \sum_{i=1}^{l} \ag_i\, \sum_{j=k+1}^{n} \ag_j\, -\, \sum_{i=1}^{k}
  \ag_i\, \sum_{j=k+1}^{n} \ag_j\, -\, \sum_{i=1}^{l} \ag_i\,
  \sum_{j=l+1}^{n} \ag_j \right]\, - }\\
 & & \\
 & & \dsp{ -\, Q_n \cdot \sum_{l=1}^{n-1}\, Q_l
  \left[\sum_{i=1}^l \ag_i\, \sum_{j=l+1}^n \ag_j \right].}
\ea
\label{16}
\ee

\nit
Performing the subtractions inside the double sum gives

\be
\ba{lll}
R_n \left[ Q_i \right] & = & \dsp{
    - \sum_{k=2}^{n-1} \sum_{l=1}^{k-1}\, Q_k \cdot
    Q_l \, \sum_{j=l+1}^k \ag_j \left[ \sum_{i=1}^l \ag_i +
    \sum_{i=k+1}^n \ag_i \right]\, -\, Q_n \cdot \sum_{l=1}^{n-1}\, Q_l\,
    \left[\sum_{i=1}^l \ag_i \sum_{j=l+1}^n \ag_j \right] }\\
  & & \\
  & = & \dsp{ - \sum_{k=2}^n \sum_{l=1}^{k-1}\, Q_k \cdot Q_l\,
     \sum_{j=l+1}^k \ag_j \lh T - \sum_{i=l+1}^k \ag_i \rh . }
\ea
\label{17}
\ee

\nit
To obtain the last line we have again used relation (\ref{11}). Note that the
remaining sums over $\ag_i$ are equal precisely the quantity $\Del_{kl}$
introduced before. Therefore eq.(\ref{17}) indeed reduces to eq.(\ref{9}).
Using
this relation under the integral with the constraints on the external momenta
and on the parameters $\ag_i$, we can then bring the $n$-particle amplitude
into
the form

\be
\ba{lll}
\Gam_n [Q_i] & = & \dsp{
   \frac{i}{n!}\, \left[ \frac{-i g}{2m (2\pi)^D} \right]^n\,
   \del^D \lh \sum_{k=1}^{n} Q_k \rh \sum_{\left\{ i_k \right\} }\,
   \int_0^{\infty} dT\, \int_0^T \prod_{i=1}^{n} d\ag_i \, \del \lh T -
   \sum_{i=1}^{n} \ag_i \rh\, \times }\\
  &  &  \\
  &  & \dsp{  \times \lh \frac{-2 \pi i m }{T} \rh^{D/2}\,
   e^{-\frac{imT}{2}\, -\, \frac{i}{m}\, \sum_{k=2}^n \sum_{l=1}^{k-1}
   \, Q_{i_k} \cdot Q_{i_l}\, G[\Del_{kl}] }. }
\ea
\label{18}
\ee

\nit
Eqs. \ref{2}, \ref{7} and \ref{18} represent three equivalent expressions for
the one-loop scalar $n$-point function, obtained from the standard Feynman
rules
for $g \vf^N$-theory. In refs.\ct{MS}-\ct{CS3} it was shown, that similar
results are obtained from the path-integral for a relativistic point-particle
of
mass $m$ calculated in a specific way. In the following we rederive some of
these results, and show that each of our three expressions results directly
from
a single functional integral using different methods of evaluation. In the
course of these derivations we will encounter yet another analytic
parametrization of the one-loop scalar integral.

\section{The path-integral representation}{\label{S3}}

The starting point for the path-integral calculations is the expression \ct{MS}

\be
\Gam_n [Q_i]\, =\, \frac{1}{n!}\, \left[\frac{-ig}{2m(2\pi)^D} \right]^n\,
   \int_0^{\infty} \frac{dT}{T}\, \oint D\xi[\tau]\, \int_0^T \prod_{i=1}^n
   d\tau_i\, e^{\frac{im}{2}\, \int_0^T d\tau \lh \dot{\xi}_{\mu}^2 - 1 \rh\,
   +\, i \sum_{j=1}^n Q_j \cdot \xi_j },
\label{19}
\ee

\nit
which is claimed to be equal to the one-loop amplitude of eq.(\ref{2}). The
right-hand side of (\ref{19}) can be interpreted as the sum over all closed
world lines (satisfying periodic boundary conditions) of a scalar particle, of
all possible proper-time lengths $T>0$, with $n$ insertions of interactions
with
an external potential, giving the particle a kick of momentum $Q_{i_k}$ at
proper time $\tau_k$ when the particle co-ordinates are $\xi^{\mu}_k =
\xi^{\mu}(\tau_k)$. Note that in pure $g \vf^N$-theory the only point-like
interaction is the one described by one single type of vertex; in particular,
the values of $\tau_k$ for two different vertices never coincide, since this
would correspond to the introduction of a new point-like vertex. This argument
can of course be made rigorous only in a renormalizable field theory;
nevertheless in the following we take for granted that all interactions are
separated in proper time.

We can now order the interactions in proper time:

\be
0 \leq \tau_1 < \tau_2 < ... < \tau_n \leq T,
\label{20}
\ee

\nit
In principle, there are $n!$ such orderings, but the various possibilities are
all
taken into account by symmetrizing over the permutations $\left\{ i_k \right\}$
of
the external momenta $\left\{ Q_{i_k} \right\}$. Therefore eq.(\ref{19}) is
equivalent to

\be
\ba{lll}
\Gam_n [Q_i] & = & \dsp{
            \frac{1}{n!}\, \left[\frac{-ig}{2m(2\pi)^D} \right]^n\, \sum_{
  \left\{ i_k \right\}}\, \int_0^{\infty} \frac{dT}{T}\, \int_0^T d\tau_n\,
  \int_0^{\tau_n} d\tau_{n-1}\, ...\, \int_0^{\tau_2} d\tau_1 \times }\\
 & & \\
 & & \dsp{ \times \oint D\xi[\tau]\, e^{\frac{im}{2}\, \int_0^T d\tau \lh
  \dot{\xi}_{\mu}^2 - 1 \rh\, +\, i \sum_{j=1}^n Q_j \cdot \xi_j }. }
\ea
\label{21}
\ee

\nit
The action integral from 0 to $T$ in the exponent can of course be decomposed
into
a sum

\be
\int_0^T d\tau = \sum_{k=1}^n\, \int_{\tau_{k-1}}^{\tau_k}\, d\tau,
\label{22}
\ee

\nit
with $\tau_0 = \tau_n$. In the following we use the standard functional
integral results for free particles in $D$-dimensional Minkowski space,
expressed by

\be
\ba{lll}
\dsp{ \int_{\xi_1}^{\xi_2} D\xi[\tau]\, e^{\frac{im}{2}\,
\int_{\tau_1}^{\tau_2}
      d\tau \lh \dot{\xi}^2_{\mu} - 1 \rh } } & = & \dsp{
      i \lh \frac{m}{2\pi i \ag} \rh^{D/2}\, e^{-\frac{im\ag}{2}\, +\,
      \frac{im}{2\ag}\, \lh \xi_2 - \xi_1 \rh^2 } }\\
 &   & \\
 & = & \dsp{ \int \frac{d^D p}{(2\pi)^D}\, e^{-\frac{i\ag}{2m}\, \lh p^2 + m^2
      \rh\, +\, i p \cdot \lh \xi_2 - \xi_1 \rh }, }
\ea
\label{23}
\ee

\nit
where $\ag = \tau_2 - \tau_1$. In eq.(\ref{21}) this result is to be used $n$
times, once for each of the functional integrals over the various sections of
the path, from $\tau_{k-1}$ to $\tau_k$. This requires the introduction of $n$
parameters

\be
\ag_1= \tau_1 - \tau_n + T, \hspace{3em} \mbox{and} \hspace{3em}
\ag_k = \tau_k - \tau_{k-1}, \hspace{1em} 2 \leq k \leq n,
\label{24}
\ee

\nit
Clearly, these parameters satisfy the constraint (\ref{11}), hence they are not
all independend; this is the result of translation invariance. The same
invariance
also makes the integral over $\tau_n$ trivial, since this amounts to an
integration over all possible choices of the origin of proper time. Thus we can
trade the multiple integral over the $\tau_k$ for a constrained integral over
the
$\ag_i$:

\be
\int_0^T d\tau_n \int_0^{\tau_{n}} d\tau_{n-1}\, ...\, \int_0^{\tau_2} d\tau_1
\, =\, T\, \int_0^T \prod_{i=1}^n d\ag_i\, \del \lh T - \sum_{j=1}^n \ag_j \rh.
\label{25}
\ee

\nit
After performing the functional integrals between each pair of successive
interaction points, and making the variable transformation (\ref{25}), one is
left only with an integral over the vertex co-ordinates $\xi^{\mu}_k$ and over
each of the momenta $p_k$ of the free particles propagating between the
vertices.
The result is

\be
\ba{lll}
\Gam_n [Q_i] & = & \dsp{ \frac{1}{n!}\, \left[ \frac{-ig}{2m (2\pi)^D}
\right]^n
   \, \sum_{\left\{ i_k \right\}}\, \int_0^{\infty} dT \int_0^T \prod_{i=1}^n
   d\ag_i\, \del \lh T - \sum_{j=1}^n \ag_j \rh\, \times } \\
  &  &  \\
  &  & \dsp{ \times \int \prod_{k=1}^n \left[ \frac{d^D \xi_k d^D p_k}{
   (2\pi)^D} \right]\, e^{-\frac{i}{2m} \sum_{l=1}^n \ag_l \lh p_l^2
   + m^2 \rh\, +\, i \sum_{l=1}^n \lh Q_{i_l} + p_l - p_{l+1} \rh \cdot
   \xi_l }. }
\ea
\label{26}
\ee

\nit
As expected, the integrations over the vertex positions $\xi_k^{\mu}$ produce
$\del$-functions for momentum conservation at each vertex. The integrals over
the momenta $p_k$ eliminate all of these $\del$-functions except one: there
remains the $\del$-function expressing conservation of the external momenta.
This leaves one momentum integral, and the formula for the amplitude becomes

\be
\ba{lll}
\Gam_n [Q_i] & = & \dsp{ \frac{1}{n!}\, \left[ \frac{-ig}{2m (2\pi)^D}
\right]^n
   \, \del^{(D)} \lh \sum_{k=1}^n Q_k \rh\, \sum_{\left\{ i_k \right\}}\,
   \int_0^{\infty} dT \int_0^T \prod_{i=1}^n d\ag_i\,
   \del \lh T - \sum_{j=1}^n \ag_j \rh\, \times } \\
  &  &  \\
  &  & \dsp{ \times \int d^D p\, e^{-\frac{i}{2m} \sum_{l=1}^n \ag_l
       \lh p_l^2 + m^2 \rh }, }
\ea
\label{27}
\ee

\nit
where the momenta $p_l$ are to be interpreted as in eq.(\ref{3}). Eq.(\ref{27})
is identical to eq.(\ref{6}). Hence from this result we can directly obtain
either eq.(\ref{2}) by integration over the parameter $T$, or eq.(\ref{7}) by
integrating out the loop momentum $p$. Therefore both these equations are
indeed
seen to follow in a straightforward way from the functional integral
(\ref{19}).

We proceed to show, that eq.(\ref{18}) can be derived from this same functional
integral, but using the evaluation procedure of ref.\ct{MS}, rather than
the procedure sketched above. Introducing the notation

\be
j^{\mu}(\tau)\, =\, \sum_{k=1}^n\, Q^{\mu}_{i_k}\, \del \lh \tau - \tau_k \rh,
\label{28}
\ee

\nit
eq.(\ref{21}) can be written in the form

\be
\ba{lll}
\Gam_n [Q_i] & = & \dsp{
  \frac{1}{n!}\, \left[\frac{-ig}{2m(2\pi)^D} \right]^n\, \sum_{
  \left\{ i_k \right\}}\, \int_0^{\infty} \frac{dT}{T}\, \int_0^T d\tau_n\,
  \int_0^{\tau_n} d\tau_{n-1}\, ...\, \int_0^{\tau_2} d\tau_1 \times }\\
 & & \\
 & & \dsp{ \times \oint D\xi[\tau]\, e^{i \int_0^T d\tau \left[-\frac{m}{2}\,
  \xi_{\mu} \lh \frac{d^2}{d\tau^2} \rh\, \xi^{\mu}\, +j \cdot \xi -
  \frac{m}{2} \right] }. }
\ea
\label{29}
\ee

\nit
Clearly the external momenta act as external forces, and the partial
integration
in the action is allowed because of the periodic boundary conditions.
Since the action integral in the exponent is a quadratic expression in the
$\xi^{\mu}(\tau)$, we can perform the functional integration by completing
the square. This requires the introduction of the inverse of the operator
$-d^2/d\tau^2$ in the space of continuous functions on a closed line of length
$T$. As discussed in ref.\ct{MS}, such an inverse does not exist on the full
function space, because of the existence of zero modes: the constant functions.
When one restricts oneself to periodic functions $\bar{\xi}(\tau)$ which have
no
constant component:

\be
\int_0^T d\tau\, \bar{\xi}(\tau)\, =\, 0,
\label{30}
\ee

\nit
the inverse does exist, and is in fact given by $G[\tau -\tau^{\prime}]$
defined
in eq.(\ref{10}):

\be
-\frac{d^2}{d\tau^2}\, G[\tau - \tau^{\prime}]\, =\,
\del \lh \tau - \tau^{\prime} \rh\, -\, \frac{1}{T}.
\label{31}
\ee

\nit
This Green function is translation invariant, reflection symmetric and periodic
with period $T$. From eq.(\ref{31}) it follows that for functions restricted
as in (\ref{30}), one has

\be
-\frac{d^2}{d\tau^2}\, \int_0^T d\tau^{\prime}\, G[\tau - \tau^{\prime}]\,
 \bar{\xi}(\tau^{\prime})\, =\,  \bar{\xi}(\tau).
\label{32}
\ee

\nit
The procedure to be followed is therefore, as explained in \ct{CS2}, to split
off
the constant mode from $\xi^{\mu}(\tau)$:

\be
\xi^{\mu}(\tau)\, =\, \xi^{\mu}_0\, +\, \bar{\xi}^{\mu}(\tau),
\label{33}
\ee

\nit
with

\be
\xi^{\mu}_0\, =\, \frac{1}{T}\, \int_0^T d\tau\, \xi^{\mu}(\tau).
\label{34}
\ee

\nit
Then the Fourier decomposition of $\bar{\xi}^{\mu}(\tau)$ has no constant
component. Defining a new variable of integration

\be
\tilde{\xi}^{\mu}(\tau)\, =\, \bar{\xi}^{\mu}(\tau)\, +\,
  \frac{1}{m}\, \int_0^T d\tau^{\prime}\, G[\tau - \tau^{\prime}]\,
  j^{\mu}(\tau^{\prime}),
\label{35}
\ee

\nit
the action integral in the exponent in (\ref{29}) can be brought into the form

\be
\ba{lll}
\lefteqn{ \dsp{
    \int_0^T d\tau\, \left[ -\frac{m}{2}\, \xi_{\mu}\, \frac{d^2}{d\tau^2}\,
    \xi^{\mu}\, +\, j_{\mu} \xi^{\mu} \right]\, =
     \int_0^T d\tau\, \left[ -\frac{m}{2}\, \tilde{\xi}_{\mu}\,
    \frac{d^2}{d\tau^2}\, \tilde{\xi}^{\mu} \right]\, - } } & & \\
  & & \\
  & & \dsp{
     -\, \frac{1}{2m}\, \int_0^T d\tau \int_0^T d\tau^{\prime}\,
    j_{\mu}(\tau)\, G[\tau - \tau^{\prime}]\, j^{\mu}(\tau^{\prime})\, +\,
    \xi^{\mu}_0\, \int_0^T d\tau\, j_{\mu}(\tau). }
\ea
\label{36}
\ee

\nit
Note that $\xi_0^{\mu}$ has become a Lagrange  multiplier imposing the
constraint

\be
\int_0^T d\tau\, j^{\mu}(\tau)\, =\,  \sum_{k=1}^n\, Q^{\mu}_{i_k}\, =\, 0
\label{37}
\ee

\nit
In the functional integral we therefore integrate separately over $\xi_0$ and
$\tilde{\xi}$; the first integration gives the $\del$-function for momentum
conservation, the second is a pure Gaussian and can be evaluated using
eq.(\ref{23}) with $\xi_2 = \xi_1$, owing to the periodic boundary conditions.
It is straightforward to see that

\be
\int_0^T d\tau \int_0^T d\tau^{\prime}\, j_{\mu}(\tau)\,
   G[\tau - \tau^{\prime}]\, j^{\mu}(\tau^{\prime})\, =\,
   \sum_{k,l=1}^n\, Q_{i_k} \cdot Q_{i_l}\, G\left[ \tau_k - \tau_l \right].
\label{38}
\ee

\nit
Therefore we obtain

\be
\ba{lll}
\Gam_n [Q_i] & = & \dsp{
  \frac{i}{n!}\, \left[\frac{-ig}{2m(2\pi)^D} \right]^n\, \del \lh
  \sum_{k=1}^n\, Q_k \rh\, \sum_{\left\{ i_k \right\}}\, \int_0^{\infty}
  \frac{dT}{T}\, \lh - \frac{2\pi i m}{T} \rh^{D/2}\,
  \times }\\
 & & \\
 & & \dsp{ \times \int_0^T d\tau_n\, \int_0^{\tau_n} d\tau_{n-1}\, ...\,
  \int_0^{\tau_2} d\tau_1\, e^{- \frac{im}{2}T\, - \frac{i}{m}\, \sum_{k=2}^n
  \sum_{l=1}^{k-1}\, Q_{i_k} \cdot Q_{i_l}\, G \left[ \tau_k - \tau_l \right]
  }. }
\ea
\label{39}
\ee

\nit
We have used the fact that $G[0] = 0$ and the reflection symmetry of
$G[\tau_k - \tau_l]$ to resum the terms in the exponent in
such a way that the argument of $G$ is always positive: $\tau_k > \tau_l$.
This accounts for a factor of 2 difference compared to eq.(\ref{38}).

Finally, using the transformation (\ref{25}), eq.(\ref{39}) is seen to equal
the right-hand side of eq.(\ref{18}). In particular, we have

\be
\tau_k\, -\, \tau_l\, =\, \sum_{j = l+1}^{k}\, \ag_j\, =\,  \Del_{kl}.
\label{40}
\ee

\nit
Therefore the two methods of evaluating the path-integral, either directly or
using the Green function technique, give fully equivalent results.

Eq.(\ref{39}) may be used to derive yet another expression for the $n$-point
scalar amplitude. Namely, if one rescales the variables $\tau_k$ like the
$\ag_k$ in (\ref{5}):

\be
u_k\, \equiv\, \frac{\tau_k}{T},
\label{41}
\ee

\nit
taking values in the range $[0,1]$, with a simultaneous rescaling
$G[\tau_k - \tau_l] \rightarrow T \tilde{G}[u_k - u_l]$, then one obtains an
integral of $\Gam$-function type, as in the right-hand side of (\ref{4}):

\be
\ba{lll}
\lefteqn{\dsp{ \Gam_n [Q_i]\, =\, \frac{i}{n!}\, \lh -2\pi i m \rh^{D/2}\,
  \left[\frac{-ig}{2m(2\pi)^D} \right]^n\, \del \lh \sum_{k=1}^n\, Q_k \rh\,
  \sum_{\left\{ i_k \right\}}\, \int_0^{\infty} dT\, T^{n-1-(D/2)}\, \times }
  }& & \\
 & & \\
 & & \dsp{ \times  \int_0^1 du_n\,
  \int_0^{u_n} du_{n-1}\, ... \, \int_0^{u_2} du_1\,
  e^{- \frac{im}{2}T\, -\, \frac{iT}{m}\, \sum_{k=2}^n \sum_{l=1}^{k-1}\,
  Q_{i_k} \cdot Q_{i_l}\, \tilde{G}[u_k - u_l] }.}
\ea
\label{42}
\ee

\nit
Remembering the $i\eps$-prescription, the integral over $T$ can be carried out
and gives

\be
\ba{lll}
\lefteqn{\dsp{ \Gam_n [Q_i]\, =\, \frac{i\pi^{D/2}}{n!}\,
  \left[\frac{-g}{(2\pi)^D} \right]^n\,  \Gam \lh n - \frac{D}{2} \rh\,
  \del \lh \sum_{k=1}^n\, Q_k \rh\, \times }} & & \\
 & & \\
 & & \dsp{ \times \sum_{\left\{ i_k \right\}}\, \int_0^1 du_n\,
  \int_0^{u_n} du_{n-1}\, ... \, \int_0^{u_2} du_1\,
  \left[ \sum_{k,l=1}^n\, Q_{i_k} \cdot Q_{i_l}\, \tilde{G}[u_k - u_l]\,
  +\, m^2 \right]^{- n + D/2}.}
\ea
\label{43}
\ee

\section{Conclusions}{\label{S4}}

{}From the computations presented here, as well as in the earlier papers
\ct{MS}-\ct{CS3}, we can draw two main conclusions. First, different ways of
evaluating the functional integral (\ref{19}) give equivalent results. Second,
these results agree fully with a direct computation of the corresponding
Feynman
diagrams, which are derived from the appropriate relativistic quantum field
theory. In comparing the various approaches we have established four different
parametrizations of the one-loop $n$-point function\footnote{An intermediate
fifth one is the highly symmetric integral expression (\ref{26}).}
given by eqs.\ \ref{2}, \ref{7}, \ref{18} and \ref{43}.

{}From these observations, it is now possible to construct directly the
effective action for the scalar field theory in the one-loop approximation.
Let $U[\vf]$ denote a potential constructed out of the scalar fields; its
momentum-space components are defined by

\be
U[\vf(\xi)]\, =\, \int \frac{d^D Q}{(2\pi)^D}\, \tilde{U}(Q)\,
                  e^{i Q \cdot \xi}.
\label{44}
\ee

\nit
Now introduce the generating functional for the one-loop scalar $n$-point
functions:

\be
\ba{lll}
W[U] & = & \dsp{ \sum_{n=0}^{\infty}\, \int \prod_{i=1}^n d^D Q_i\:
           \Gam_n [Q_i]\, \tilde{U}(Q_n)\, ...\, \tilde{U}(Q_1) } \\
  &   & \\
  & = & \dsp{ \sum_{n=0}^{\infty}\, \frac{1}{n!}\, \left[ \frac{-ig}{2m}
        \right]^n\, \int_0^{\infty} \frac{dT}{T}\, \oint D\xi [\tau]\,
        \prod_{i=1}^n\, \lh \int_0^T  d\tau_i\, U[\vf(\xi_i)] \rh\,
        e^{\frac{im}{2}\, \int_0^T d\tau\, \lh \dot{\xi}_{\mu}^2 - 1 \rh} .}
\ea
\label{45}
\ee

\nit
In this expression the sum over $n$ can be carried out to give an exponential,
leading to the result

\be
W[U]\, =\, \int_0^{\infty} \frac{dT}{T}\, \oint D\xi [\tau]\,
           e^{i \int_0^T d\tau\, \lh \frac{m}{2}\, \dot{\xi}^2_{\mu}\, -\,
           \frac{m}{2}\, -\, \frac{g}{2m}\, U[\vf(\xi)] \rh }.
\label{46}
\ee

\nit
This is the path-integral for a relativistic scalar particle moving in an
external scalar potential $U[\vf]$. This path-integral is therefore seen to
generate by its expansion in powers of $g$ all one-loop $n$-point functions for
the scalar field theory for which $U[\vf]$ is the second derivative of the
classical potential. For example, the one-loop scalar integrals of
$g\vf^4$-theory in four dimensions are obtained by taking $U[\vf] = \vf^2/2$.
$W[U]$ thus corresponds to the one-loop approximation to the logarithm of
the generating functional for Green functions of the corresponding field theory
(modulo a constant):

\be
W[U]\, =\, - \log \det \lh - \Box + m^2 + g U[\vf] \rh .
\label{47}
\ee

\nit
In this way we reproduce, here for pure scalar field theories, the starting
point of ref.\ct{MS} and the subsequent papers cited below.
\vspace{7ex}

\nit
{\bf Acknowledgement}
\vspace{2ex}

\nit
The work described here is supported in part by the Human Capital and Mobility
Program through the network on {\em Constrained Dynamical Systems}.

\end{document}